\documentclass[11pt,nofootinbib,showpacs,notitlepage]{revtex4-1}
\usepackage[T1]{fontenc}
\usepackage{babel}
\usepackage[utf8]{inputenc}
\usepackage{babel,amsmath,amssymb,mathrsfs,amsfonts,calligra,euscript,textcomp,graphicx,wrapfig,fontenc,pst-blur,tabu,lipsum,%
float,empheq,pst-node,environ,framed,ccaption,capt-of,enumitem,subcaption,enumitem,framed,txfonts,lmodern,cancel,diagbox,multirow,float,tabularx}
\usepackage[section]{placeins}
\definecolor{darkgreen}{rgb}{0,.7,0}
\usepackage[colorlinks,linkcolor=blue,citecolor=green,pagebackref=true]{hyperref}
\def\dac{\displaystyle\frac}

\def\[{\left[}
\def\]{\right]}
\def\({\left(}
\def\){\right)}

\newcommand{\const}{\mathop{\rm const}\nolimits}
\newcommand{\E}{\mathop{\rm e}\nolimits}
\newcommand{\m}{\mathop{\mathcal{M}}\nolimits}

\newcommand{\eq}[1]{\begin{equation}#1\end{equation}}
\newcommand{\eqs}[1]{\begin{equation*}#1\end{equation*}}

\newcommand{\bgp}[1]{\bigl(#1\bigr)}
\newcommand{\lrp}[1]{\left(#1\right)}

\newcommand{\lrbr}[1]{\left\{#1\right\}}

\begin{document}

\baselineskip7mm

\title{Some aspects of the cosmological dynamics in Einstein-Gauss-Bonnet gravity}

\author{Dmitry Chirkov}
\affiliation{Sternberg Astronomical Institute, Moscow State University, Moscow, Russia}
\affiliation{Bauman Moscow State Technical University, Moscow, Russia}
\author{Sergey A. Pavluchenko}
\affiliation{Programa de P\'os-Gradua\noexpand\c c\~ao em F\'isica, Universidade Federal do Maranh\~ao (UFMA), 65085-580, S\~ao Lu\'is, Maranh\~ao, Brazil}

\begin{abstract}
We study some aspects of dynamical compactification scenario where stabilisation of extra dimensions occurs due to presence the Gauss-Bonnet term and non-zero spatial curvature. In the framework of the model under consideration there exists two-stages scenario of evolution of a Universe: on the first stage, the space evolves from a totally anisotropic state to the state with 3-dimensional (corresponding to our ``real''\, world) expanding and $D$-dimensional contracting isotropic subspaces; on the second stage, constant curvature of extra dimensions begins to play role and provide compactification of extra dimensions. It is already known that such a scenario is realizable when constant curvature of extra dimensions is negative. Here we show that a range of coupling constants for which exponential solutions with 3-dimensional expanding and $D$-dimensional contracting isotropic subspaces are stable is located in a zone where compactification solutions with positively curved extra space are unstable, so that two-stage scenario analogous to the one described above is \emph{not} realizable. Also we study ``nearly-Friedmann''\, regime for the case of arbitrary constant curvature of extra dimensions and describe
new parametrization of the general solution for the model under consideration which provide elegant way of describing areas of existence over parameters space.
\end{abstract}

\maketitle

\section{Introduction}
General Relativity (GR) remains the gold standard in describing the behaviour of black holes, the propagation of gravitational waves, the formation of all structures in the Universe etc. However it has still some limits too. One of the issues of General Relativity is that in order to fit the astrophysical observations one must assume the existence of dark matter and dark energy. In other words, if General Relativity is correct, it seems that around 96\% of the Universe should be in the form of energy densities that do not interact electromagnetically.

This problem led to the idea of modifying gravity on cosmological scales. There are a lot of ways to modify theory of gravity. Since the early days after Einstein’s publication of his theory there were proposals being made on how to incorporate it in a more unified theory. It is appropriate to mention here Eddington’s theory of connections~\cite{Eddington}, the higher dimensional theories of Kaluza and Klein~\cite{KK}, Weyl’s scale independent theory~\cite{Weyl}.

Building on the works of Weyl, Sakharov proposed that the Einstein-Hilbert action is a first approximation to a much more complicated action and, in part, fluctuations in space-time itself lead to higher powers corrections to Einstein’s theory~\cite{Mod-grav-cosmol}. Later Stelle and Starobinsky developed Sakharov’s approach: Stelle showed that these theories are renormalizable in the presence of matter fields at the one loop level~\cite{Stelle}; Starobinsky found cosmological consequences of these theories~\cite{Starobinsky}.

A very popular modified theory of gravity used in order to explain the accelerated expansion of the universe is $f(R)$ gravity~\cite{Sotiriou-Faraoni}. In the framework of $f(R)$ gravity dynamical equations are 4th order in derivatives instead of the 2nd order for GR; this causes appearance nonphysical solutions like ``false radiation''\, vacuum isotropic solution~\cite{Muller}.

One of the theories which keeps the order of equations of motion the same as in GR is Lovelock gravity~\cite{Lovelock}; moreover, in the standard metric formalism the Lovelock gravity is the only generalisation of GR with this property. Lovelock gravity is a natural generalization of GR in the following sense: it is well known that the Einstein tensor is the only symmetric and divergence free tensor depending only on the metric and its first and second derivatives (with a linear dependence on second derivatives); dropping the condition of linear dependence on second derivatives, we obtain the Lovelock tensor -- the most general tensor which satisfies other mentioned conditions.

Lovelock gravity gives corrections to GR only in higher-dimensional spacetime, so it is natural to consider it in the framework of multidimensional cosmology. Extra dimensions, at least macroscopically, can not be observed, so it is necessary to suppose that they are compactified to a very small scale. In order to our theoretical model fit to ``real''\, world we need to require that three dimensions are expanding isotropically while extra dimensions are contracting (or have been contracting during some period of Universe evolution). A lot of studies start from decomposition of manifold under consideration into a product of 3-dimensional ``external''\, and $D$-dimensional ``internal''\, spaces without any attempt to clarify its 
origin~\cite{add1, Muller-Hoissen-1,Muller-Hoissen-2,Deruelle}.
One of the first exact static solutions with metric being a cross product of a (3+1)-dimensional Minkowski manifold times a constant curvature ``inner space''\, were found in~\cite{Muller-Hoissen-1}; later Minkowski manifold was replaced with a constant curvature Lorentzian manifold in~\cite{Deruelle}. Model with Friedman-Robertson-Walker manifold as (3+1)-dimensional section and with constant sized extra dimensions was considered in~\cite{Muller-Hoissen-2}; there it was explicitly showed that to have more realistic model one needs to consider the dynamical evolution of the extra dimensional scale factor. An attempt to realise dynamical compactification scenario was done in~\cite{Ishihara} where both the (3+1)-dimensional and the extra dimensional scale factors are exponential functions.

In this paper we consider quadratic Lovelock theory also known as Einstein-Gauss-Bonnet (EGB) gravity~\cite{Garraffo}. The Lagrangian of this theory is the sum of three terms: $\Lambda$-term
(formally, it is boundary term -- see Section~\ref{sec.fried} for details), the Einstein-Hilbert term and so-called Gauss-Bonnet (GB) term which is quadratic in the curvature and reads $R_{\mu\nu\rho\sigma}R^{\mu\nu\rho\sigma}-4R_{\mu\nu}R^{\mu\nu}+R^2$. For (3+1)-dimensional space-time the Gauss-Bonnet term is topological and does not affect the dynamical equations; in dimension higher than four this term gives a non-trivial contribution to equation of motion.

We study solutions with exponential time behavior of scale factors (i.e. constant Hubble parameters). Exponential solutions are less studied but due to their ability to compactify extra dimensions fast and reliably has a great potential. This class of solutions have been initially found in EGB theory in the regime when the Gauss-Bonnet term is dominated~\cite{Ivashchuk} and later have been generalized to full EGB theory~\cite{DPT-15, DPT-14, CST2}; the general scheme for constructing exponential solutions in general Lovelock gravity of any order and in any dimensions was developed in~\cite{DPT-15}.

It worth mentioning that apart from the exponential solutions, for cosmology of importance power-law solutions as well: exponential solutions resemble de Sitter stage while power-law are analogues 
of Friedmann regime. Power-law solutions in EGB cosmology also were studied intensively~\cite{PK1, PK2}; let us also note that in~\cite{PT} links between power-law and 
exponential solutions in GB and EGB cosmologies were studied. Additionally, in current paper we study model with nonzero $\Lambda$-term but formally one could introduce other sources of matter, say,
perfect fluid and cosmological models with such sources also were studied in the framework of GB and EGB cosmology as well~\cite{PK2, CST2, prd10, my18d}.

Analysis of exponential solutions in EGB gravity demonstrated that they exists only if the space has isotropic subspaces~\cite{DPT-14}; this fact remains valid also for general Lovelock model~\cite{DPT-15}. It should be emphasized that this splitting into isotropic subspaces appears naturally from equations of motion as a condition for such solutions to exist. Detailed study of this situation has exposed that there are solutions where three dimensions (corresponding to the ``real world'') are expanding, while remaining dimensions are contracting making the compactification viable. Stability of such solutions have been studied in~\cite{stab,Ivas-16}.

Realistic compactification scenario implies that the 3-dimensional Hubble parameter and the extra dimensional scale factor tend to constant values asymptotically. 
In standard cosmology spatial curvature plays an important role, say, it could change asymptotics for inflationary cosmology~\cite{infl}, so it is natural to assume that it also affects
cosmological dynamics in EGB gravity and study its effects.
In~\cite{CGPT1} it was found that such scenario is realised in the case that the curvature of the extra dimensions is negative and the Einstein-Gauss-Bonnet theory does not admit a maximally symmetric solution. In~\cite{CGPT2} a detailed analysis of the cosmological dynamics in this
model with generic couplings was performed; later in~\cite{CGPT3} it was demonstrated that Friedmann late-time dynamics in three-dimensional part could be restored assuming additional constraint on couplings.

Building on these works, in~\cite{CGT-2020} the model that consists of two stages was developed. On the first stage, the space evolves from a totally anisotropic state to the state with 3-dimensional expanding and $D$-dimensional contracting isotropic subspaces (apart from~\cite{CGT-2020} first stage alone was numerically considered in~\cite{PT2017}); 
on the second stage, constant negative curvature of extra dimensions begins to play the role and provide compactification of these extra dimensions.

In the current paper we address two problems: a possibility of constructing two-stages scenario for the case of constant positive curvature of extra dimensions as well as restoring Friedmann late-time dynamics in three-dimensional subspace for the case of arbitrary constant curvature of extra dimensions. Additionally, we present nice graphical representation of the general solution with one 
spatially curved and another flat manifolds. 

The structure of the manuscript is the following: in the next section the EGB action and the dynamical equations are presented;  after that we present 
 stability of exponential solutions with 3-dimensional expanding and $D$-dimensional contracting isotropic subspaces; then we present some properties of the general solution which were left behind
 in our previous study~\cite{we2020} including new parametrization of the solution with nice
graphical meaning;
 the fifth section is devoted to the problem of recovering ``standard''\, Friedmann dynamics in $(3+1)$-dimensional subspace; in the last section the conclusions will be given.

\section{The set-up}
Let us consider $(D+4)$-dimensional spacetime $\mathcal{M}$ with the Einstein-Gauss-Bonnet action
\eq{S=\int_{\mathcal{M}}d^{D+4}x\sqrt{|g|}\left\{R-\Lambda+\alpha\bgp{R_{\mu\nu\rho\sigma}R^{\mu\nu\rho\sigma}-4R_{\mu\nu}R^{\mu\nu}+R^2}\right\},\label{Sorig}}
where $|g|$ is the determinant of metric tensor; $\Lambda$ is the cosmological term (again, it is rather boundary term -- see Section~\ref{sec.fried} below for details); 
$R,R_{\alpha\beta},R_{\alpha\beta\gamma\delta}$ are the $(D+4)$-dimensional scalar curvature, Ricci tensor and Riemann tensor respectively; $\alpha$ is the coupling constant. Here and after Greek indices run from 0 to $D+4$, Latin indices from 1 to $D+4$ unless otherwise stated.

We assume that the space-time $\mathcal{M}$ is a warped product $\m_4\times \m_D$ where $\m_4$ is a flat Friedman-Robertson-Walker manifold with scale factor $a(t)$, $\m_D$ is a $D$-dimensional Euclidean compact and constant curvature manifold with scale factor $b(t)$ and spatial curvature $\gamma_{D}$ . We choose the ansatz for the metric in the form
\eq{ds^2=-dt^2+a(t)^2d\Sigma^2_{3}+b(t)^2d\Sigma^2_{D}\label{metric}}
We denote $H=\frac{a'}{a}$; then $\frac{a''}{a}=H'+H^2$. Equations of motion~\cite{CGPT1, CGPT2, CGT-2020} reads
\eq{\begin{split}
       &\frac{6}{D+1}\left(\frac{2H b'(D+1)!}{b (D-1)!}+\frac{H^2(D+1)!}{D!}+\frac{(\gamma_{D}+b'^2)(D+1)!}{2b^2(D-2)!}+\frac{b''(D+1)!}{b(D-1)!}+\frac{2(H'+H^2)(D+1)!}{D!}\right)+ \\
         &+6D\alpha\Biggl(\frac{(\gamma_{D}+b'^2)^2(D-1)!}{2b^4(D-4)!}+
      \frac{8b''b'H(D-1)!}{b^2(D-2)!}+\frac{4(\gamma_{D}+b'^2)(H'+H^2)(D-1)!}{b^2(D-2)!}+\\
         &\hspace{1.5cm} +\frac{4H^2b''}{b}+\frac{4Hb'(\gamma_{D}+b'^2)(D-1)!}{b^3(D-3)!}+\frac{4H^2b'^2(D-1)!}{b^2(D-2)!}+\\
         &\hspace{1.5cm} +\frac{2H^2(\gamma_{D}+b'^2)(D-1)!}{b^2(D-2)!}+\frac{8(H'+H^2)Hb'}{b}+\frac{2b''(\gamma_{D}+b'^2)(D-1)!}{b^3(D-3)!}\Biggr)-3\Lambda=0
    \end{split}
\label{E1}}
\eq{\begin{split}
       &\frac{6}{D+1}\left(\frac{H b'(D+1)!}{b (D-2)!}+\frac{H^2(D+1)!}{(D-1)!}+\frac{(\gamma_{D}+b'^2)(D+1)!}{6b^2(D-3)!}+\frac{b''(D+1)!}{3b(D-2)!}+\frac{(H'+H^2)(D+1)!}{(D-1)!}\right)+ \\
         &+6D\alpha\Biggl(\frac{(\gamma_{D}+b'^2)^2(D-1)!}{6b^4(D-5)!}+\frac{2H^2(\gamma_{D}+b'^2)(D-1)!}{b^2(D-3)!}+
      \frac{4b''b'H(D-1)!}{b^2(D-3)!}+\frac{4H^3b'(D-1)!}{b(D-2)!}+\\
         & \hspace{1.5cm}+\frac{4H^2b''(D-1)!}{b(D-2)!}+\frac{2Hb'(\gamma_{D}+b'^2)(D-1)!}{b^3(D-4)!}+\frac{4H^2b'^2(D-1)!}{b^2(D-3)!}+
         \frac{8(H'+H^2)Hb'(D-1)!}{b(D-2)!}+ \\
         &\hspace{1.5cm}+\frac{2b''(\gamma_{D}+b'^2)(D-1)!}{3b^3(D-4)!}+\frac{2(H'+H^2)(\gamma_{D}+b'^2)(D-1)!}{b^2(D-3)!}+4H^2(H'+H^2)\Biggr)-D\Lambda=0
    \end{split}
\label{E2}}
and constraint
\eq{\begin{split}
       &\frac{6}{D+1}\left(\frac{H b'(D+1)!}{b (D-1)!}+\frac{H^2(D+1)!}{D!}+\frac{(\gamma_{D}+b'^2)(D+1)!}{6b^2(D-2)!}\right)+
         6D\alpha\Biggl(\frac{(\gamma_{D}+b'^2)^2(D-1)!}{6b^4(D-4)!}+\\
      &\hspace{1.5cm}+\frac{2H^2(\gamma_{D}+b'^2)(D-1)!}{b^2(D-2)!}+\frac{4H^3b'}{b}+\frac{4H^2b'^2(D-1)!}{b^2(D-2)!}+
      \frac{2Hb'(\gamma_{D}+b'^2)(D-1)!}{b^3(D-3)!}\Biggr)-\Lambda=0.
    \end{split}
\label{E0}}

\section{Stability of exponential solutions with 3-dimensional expanding and $D$-dimensional contracting isotropic subspaces}
Let us consider flat anisotropic $(D+4)$-dimensional spacetime with the metric ansatz of the form
\eq{ds^2=-dt^2+\sum\limits_{k=1}^{D+4}a^2_k(t)dx_k^2}
where $a_1(t),\ldots,a_D(t)$ are scale factors. In spatially flat model scale factors $a_k(t)$ are defined only up to a constant factor, so it makes sense use the Hubble parameters $H_k=\frac{\dot{a}_k}{a_k},\;k=1,\ldots,D$ instead of scale factors. With this in mind we obtain $D+3$ equations of motions
\eq{\sum\limits_{i\ne j}H_i^2+\sum\limits_{\{i>k\}\ne j}H_i H_k+4\alpha\sum\limits_{i\ne j}H_i^2\sum\limits_{\{k>l\}\ne\{i, j\}}H_k H_l+12\alpha\sum\limits_{\{i>k>l>m\}\ne j}H_i H_k H_l H_m=\Lambda\label{eq.of.motion}}
and constraint
\eq{\sum\limits_{i>j}H_i H_j+12\alpha\sum\limits_{i>j>k>l}H_i H_j H_k H_l=\Lambda\label{constraint}}

In this section we consider spaces with 3-dimensional expanding and $D$-dimensional contracting subspaces, so without loss of generality we can set $H_1=H_2=H_3\equiv H>0,\;H_4=\ldots=H_{D+3}\equiv h<0$; in this case the system of equations~(\ref{eq.of.motion})--(\ref{constraint}) takes the form:
\eq{\begin{array}{c}
      3H^2+2DHh+\frac{D(D+1)}{2}h^2+ \\
      +4\alpha\lrbr{2DH^3h+\frac{D(5D-3)}{2}H^2h^2+D^2(D-1)Hh^3+\frac{(D-2)(D-1)D(D+1)}{8}}=\Lambda
    \end{array}
\label{eqH1}}
\eq{\begin{array}{c}
      6H^2+3(D-1)Hh+\frac{(D-1)D}{2}h^2++12\alpha\lrbr{H^4+\right. \\
      \left.+3(D-1)H^3h+(D-1)(2D-3)H^2h^2+\frac{(D-2)(D-1)^2}{2}Hh^3+\frac{(D-3)(D-2)(D-1)D}{24}h^4}=\Lambda
    \end{array}
\label{eqHD}}
\eq{\begin{array}{c}
      3H^2+3DHh+\frac{(D-1)D}{2}h^2+ \\
      +12\alpha\lrbr{DH^3h+\frac{3(D-1)D}{2}H^2h^2+\frac{(D-2)(D-1)D}{2}Hh^3+\frac{(D-3)(D-2)(D-1)D}{24}h^4}=\Lambda
    \end{array}
\label{C}}
Instead of equations~(\ref{eqH1})--(\ref{C}) we use equivalent system:
\eqs{\left\{\begin{array}{c}
              (\ref{eqH1}) \\
              (\ref{eqHD}) \\
              (\ref{C})
            \end{array}\right.\Longleftrightarrow
            \left\{\begin{array}{l}
              (\ref{eqH1})-(\ref{C}) \\
              (\ref{eqHD})-(\ref{C}) \\
              (\ref{eqHD})
            \end{array}\right.\Longleftrightarrow
            \left\{\begin{array}{l}
              Dh(h-H)Q=0\qquad\qquad (*) \\
              (D-1)H(H-h)Q=0\quad (**) \\
              (\ref{eqHD})
            \end{array}\right.
}
where $Q=1+4\alpha\lrp{H^2+2(D-1)Hh+\frac{(D-1)(D-2)}{2}h^2}$. Since we assume that $h\ne0,\;H\ne0,\;h\ne H$ it follows from equations $(*)$ and $(**)$ that
\eq{Q=0\Longleftrightarrow 1+4\alpha\lrp{H^2+2(D-1)Hh+\frac{(D-1)(D-2)}{2}h^2}=0\label{Q}}
Let us simplify equation~(\ref{eqHD}). Expressing $\alpha$ from~(\ref{Q}) and substituting it into~(\ref{eqHD}) we get
\eq{3H^4+6(D-1)H^3h+\frac{(7D-6)(D-1)H^2h^2}{2}+(D-1)^2DHh^3+\frac{(D-2)(D-1)D(D+1)h^4}{8}=-\frac{\xi}{4\alpha^2}\label{eqHD-transformed}}
where $\xi=\Lambda\alpha$. So, solutions which are the warped product of 3-dimensional expanding and $D$-dimensional contracting isotropic subspaces obey the equations~(\ref{Q})--(\ref{eqHD-transformed}). Solving~(\ref{Q}) with respect to $H$ we obtain
\eq{H_-=\frac {-2\left(D-1\right)\alpha h-\sqrt{2\alpha^2h^2D(D-1)-\alpha}}{2\alpha},\quad H_+=\frac {-2\left(D-1\right)\alpha h+\sqrt{2\alpha^2h^2D(D-1)-\alpha}}{2\alpha}}
We already know~\cite{we2020} that solutions with $\gamma_D>0$ exist and stable only for $\alpha<0$, so in what follows (to the end of this section) we consider only negative values of $\alpha$.

It was proved in~\cite{Ivas-16} that exponential solution $ds^2=-dt^2+\sum\limits_{k}\E^{2H_kt}dx_k^2,\;H_k=\const$ is stable under perturbations
$H_k+\delta H_k(t)$ as $t \to + \infty$ if $\sum\limits_{k}H_k>0$. Applied to our case it implies that a solution to equations~(\ref{Q})--(\ref{eqHD-transformed}) is stable if $3H+Dh>0$. It is easy to check that $3H_++Dh<0$ for all $h<0, \alpha<0$ where as $3H_-+Dh>0$ for all $h<0, \alpha<0$. Therefore only solutions with $H_-$ are stable.

Substituting $H_-$ into~(\ref{eqHD-transformed}) and solving the resulting equation with respect to $\xi$ we obtain
\eq{\xi(y)=-\frac{3}{2}+6y(D-1)\lrp{2D(D-1)y^2+1}^{3/2}-D(D-1)(17D^2-25D+6)y^4-(D-1)(13D-6)y^2}
where $y=h|\alpha|,\;y<0$ since we consider $h<0$. It is easy to check that the function $\xi(y)$ increase monotonically for $h<0,\;D\geqslant2$ and $\xi(0)=-\frac{3}{2}$; it means that $\xi<-\frac{3}{2}$ for all $h<0,\;D\geqslant2$.

So, we can conclude that exponential solutions with 3 expanding and $D$ contracting subspaces are stable only for $\xi<-\frac{3}{2}$ where as stable compactified solutions in $(D+4)$ dimensions with $\gamma_D>0$ are stable only for $\xi>-\frac{3}{2}$, therefore we can not realise 2-stage evolution scheme for the case of positive curvature $\gamma_D$ as we have done it for the case of negative curvature~\cite{CGT-2020}.

\section{Case with asymptotically static curved extra dimensions}

Solutions with asymptotically static curved extra dimensions were studied in~\cite{CGPT1, CGPT2, CGPT3} (the case with negative spatial curvature) and recently in~\cite{we2020} (general case). 
In~\cite{CGPT1} we demonstrated that if the spatial section could be viewed as a product of flat three-dimensional and curved extra-dimensional (with negative spatial curvature) subspaces
then there exist a regime when the expansion rate of three-dimensional subspace (``our Universe'') approach constant rate and the scale factor corresponding to extra dimensions (``size of extra
 dimensions'') also reach constant. So that we naturally end up with a cosmological model with constant Hubble parameter in ``our Universe'' with constant-sized extra dimensions. This model was
 further studied in~\cite{CGPT2, CGPT3} and recently in~\cite{we2020} we extended analysis to positive curvature of extra dimensions as well. 

General solution for (\ref{E1})--(\ref{E0}) under asymptotically static curved extra dimensions conditions ($H = H_0$, $b = b_0$)
is given in~\cite{we2020} and it has 1-parametric closed form

\begin{equation}
\begin{array}{l}
\xi = \dac{12z(1+2z)\zeta^2 + \gamma_D(D-1)(D-2)(24z+1)\zeta + \gamma_D^2(D-1)(D-2)(D-3)(D-4)}{\zeta^2}, \\ \\
\zeta_\pm = \dac{-\gamma_D (D-1) (6Dz - 24z - 1 \pm \sqrt{\mathcal{D}}) }{6z (4z+1)},~\mbox{where} \\ \\
\mathcal{D} = (D-1) \( (D-1)\gamma_D^2 (6Dz - 24z - 1)^2 + 24z(D-2)(D-3)(4z+1)   \), \\ \\
\xi = \alpha\Lambda,~ \zeta = b_0^2/\alpha,~ z = \alpha H_0^2.
\end{array} \label{3pD_zeta}
\end{equation}

\begin{figure}
\includegraphics[width=0.8\textwidth, angle=0]{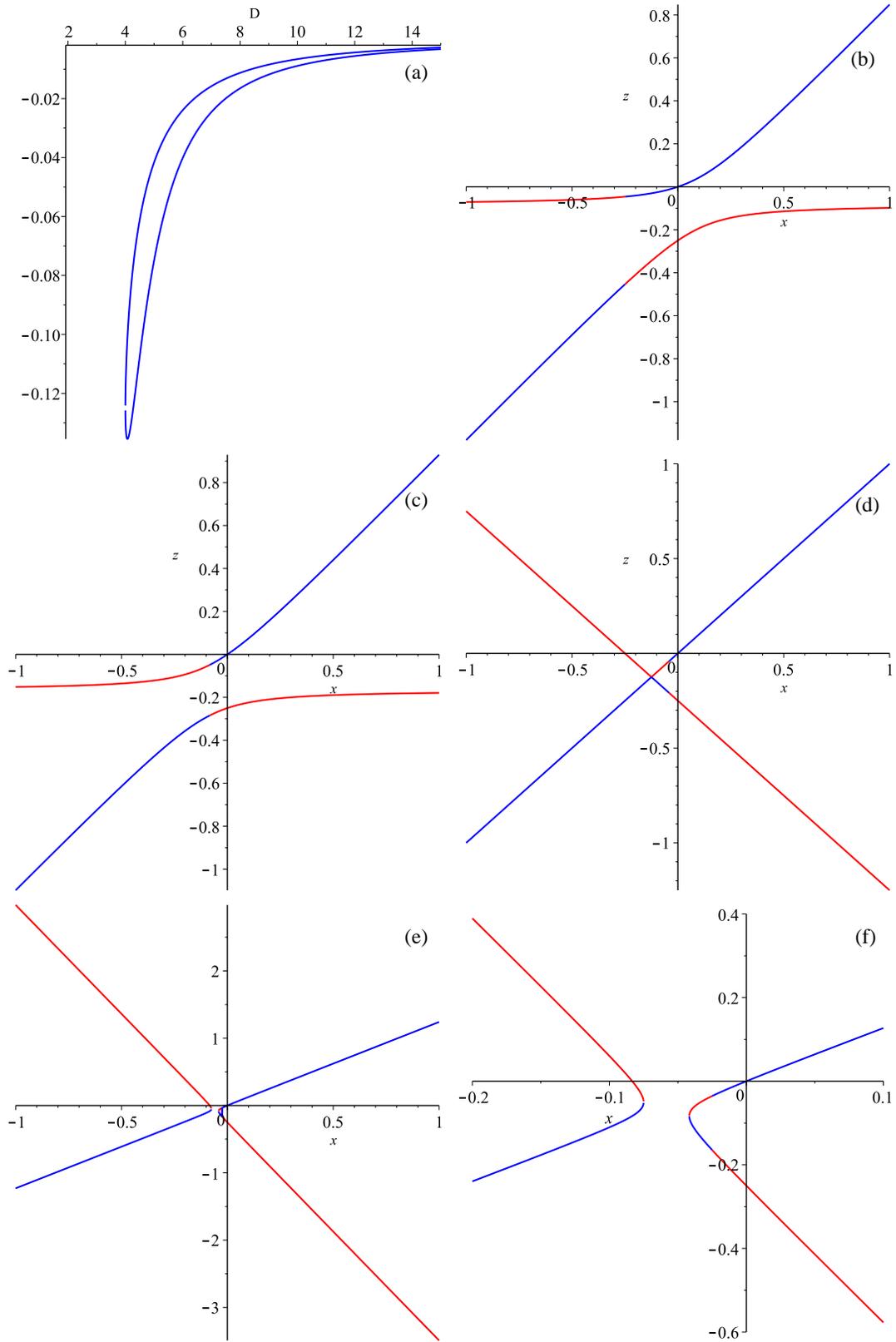}
\caption{Visualization of general behavior for all $D$: excluded range for $x$ in (a) panel, solution in $\{x,\,z\}$ coordinates for $D=2$ in (b), $D=3$ in (c), $D=4$ in (d) and general $D>4$ case
in (e) and (f) panels (see the text for more details).}\label{fig_xz}
\end{figure}

The solution and its properties were studied in~\cite{we2020} but there are two details that were left behind and we want to stress on. If we solve the system (\ref{E1})--(\ref{E0}) 
with respect to $z$ instead of $\xi$ and $\zeta$, we would obtain 
\eq{z=\frac {-D\left(D-1\right)\left(D-2\right)\left(D-3\right)x^2-\left(D-1\right)Dx+\xi}{6+12\left(D-1\right)Dx}\label{eq:z},}
where $x=\alpha\gamma_D/b_0^2$. We can note from (\ref{eq:z}) that for $D\geqslant 4$ the numerator becomes quadratic equation with
discriminant $\mathcal{D}=4(D-2)(D-3)\xi + D(D-1)$, so that for $\xi > -D(D-1)/(4(D-2)(D-3))$ the discriminant is positive and there exist ``forbidden'' range for $x$. This is happening for the
following reason: from (\ref{3pD_zeta}) it is clear that sign of $z$ should coincide with sign of $\alpha$; when the discriminant is negative, numerator of (\ref{eq:z}) is negative (since the 
multiplier before $x^2$ is negative) and by choosing $\alpha < 0$ signs for $z$ and $\alpha$ always coincide. But if the discriminant is positive, numerator swaps signs and choosing $\alpha$
is not working everywhere -- so that there is a range where signs are opposite and this is exactly the range in question. As of the denominator, one can demonstrate that if $x_\pm$ are roots of
the numerator and $x_{sep}$ is root of denominator, then $x_- < x_+ < x_{sep} < 0$, so that both $x_\pm$ (and so the entire range in question) always lies within $x_\pm < x_{sep}$ so that for
our cause the denominator is always positive. 
Let us also note that this condition for $\xi$ exactly coincide with derived earlier in ~\cite{we2020} condition for co-existence of compactified and isotropic solutions. 
In Fig.~\ref{fig_xz}(a) we presented this range for different $D$.

From definition of $z=\alpha H_0^2$ we can see that $\alpha$ and $z$ always have the same sign. But if we additionally use definition $x=\alpha\gamma_D/b_0^2$, we can notice that signs 
for $\{x,\,z\}$ correspond to
signs for $\{\alpha,\,\gamma_D\}$: for $\alpha > 0$ we have $x>0$ if $\gamma_D>0$ and $x<0$ if $\gamma_D<0$; for $\alpha < 0$ we have $x<0$ if $\gamma_D>0$ and $x>0$ if $\gamma_D<0$. So that we can plot
all possible cases for all sign combination of $\alpha$ and $\gamma_D$ on a single plot in $\{x,\,z\}$ plane. To do this, we substitute two branches of $\xi_{1,\,2}$ 
($\zeta_\pm$ substituted into $\xi$ -- both given in (\ref{3pD_zeta})) into (\ref{eq:z}) 
to obtain two branches $z_{1,\,2} = z_{1,\,2} (x, D)$. We plot them in Figs.~\ref{fig_xz}(b)--(f): $D=2$ case in (b) panel, $D=3$ case in (c) panel, $D=4$ case in (d) panel and general $D>4$
case in (e) and (f) panels: general view in (e) and focus on the gap in (f). Case which is plotted in Fig.~\ref{fig_xz}(e, f) is $D=5$; for higher $D$ the size of the gap becomes smaller and located closer to zero -- according to Fig.~\ref{fig_xz}(a).

Looking at Fig.~\ref{fig_xz} we immediately can see the difference between $D=2,\,3$ cases, $D=4$ and general $D>4$ cases: $D=2,\,3$ cases (see Figs.~\ref{fig_xz}(b, c)) ``avoid'' second quadrant,
meaning that there are no solutions for $\{z>0,\,x<0\} \leftrightarrow \{\alpha > 0,\,\gamma_D<0\}$ -- and that is exactly what we derived in~\cite{we2020}. Cases with $D\geqslant 4$ has solutions 
in all quadrants
but for $D=4$ there is no ``forbidden'' range for $x$ (compare Fig.~\ref{fig_xz}(d) with Figs.~\ref{fig_xz}(e, f)).

\section{``Nearly-Friedmann'' regime}
\label{sec.fried}

As we already mentioned, the main reason behind our study of EGB models is the description of successful compactification, as it is the only ``bridge'' which connects this theory with reality. After
describing  solutions with compactification, next step would be attempt to recover ``standard'' Friedmann dynamics in $(3+1)$-dimensional subspace. 
Similar procedure was done in~\cite{CGPT3} for $\gamma_D=-1$ case
and now we extend it to the general $\gamma_D$ case. Asymptotically-Friedmann regime would have $b(t)=b_0$ (as the size of extra dimensions already reached ``stable'' value) and $\dot H = H = 0$
(as it is low-energy regime -- at least, low-energy with respect to high-energy ``pure'' Gauss-Bonnet regime). Substituting these conditions into the equations of motion~(\ref{E1})--(\ref{E0}), we end up with two equations, linking together parameters of the theory:

\begin{equation}
\begin{array}{l}
\dac{\gamma_D D(D-1)}{b_0^2} + \dac{\alpha\gamma_D^2 D(D-1)(D-2)(D-3)}{b_0^4} = \Lambda, \\ \\
\dac{\gamma_D (D-1)(D-2)}{b_0^2} + \dac{\alpha\gamma_D^2 (D-1)(D-2)(D-3)(D-4)}{b_0^4} = \Lambda.
\end{array} \label{efL_1}
\end{equation}

Solving this system gives us

\begin{equation}
\begin{array}{l}
\dac{\alpha\gamma_D}{b_0^2} = - \dac{1}{2(D-2)(D-3)},~~\xi=\alpha\Lambda = - \dac{D(D-1)}{4(D-2)(D-3)}
\end{array} \label{efL_2}
\end{equation}

\noindent or, cancelling $\alpha$ in both equations,

\begin{equation}
\begin{array}{l}
\Lambda = \dac{\gamma_D}{2b_0^2}.
\end{array} \label{efL_3}
\end{equation}

Analysing (\ref{efL_2})--(\ref{efL_3}) we can notice that $\alpha$ and $\Lambda$ as well as $\alpha$ and $\gamma_D$ always have opposite sign while $\gamma_D$ and $\Lambda$ have the same sign.

We can rewrite the constraint equation (\ref{E0}) in the following form

\begin{equation}
\begin{array}{l}
G_{N, \mbox{eff}} H^2 = \Lambda_{4D, \mbox{eff}},
\end{array} \label{efL_4}
\end{equation}

\noindent where

\begin{equation}
\begin{array}{l}
G_{N, \mbox{eff}} = 6 \(  1+\dac{2\alpha\gamma_D D(D-1)}{b_0^2} \),~~\Lambda_{4D, \mbox{eff}} = \Lambda + \dac{\gamma_D D(D-1)}{b_0^2} + \dac{\alpha\gamma_D^2 D(D-1)(D-2)(D-3)}{b_0^4}.
\end{array} \label{efL_5}
\end{equation}

Now if we substitute (\ref{efL_2}) into (\ref{efL_5}), we obtain

\begin{equation}
\begin{array}{l}
G_{N, \mbox{eff}} = 6 \(  1 - \dac{D(D-1)}{(D-2)(D-3)} \) < 0,~~\Lambda_{4D, \mbox{eff}} = - \dac{D(D-1)}{2(D-2)(D-3)} <0,
\end{array} \label{efL_6}
\end{equation}

\noindent so that swapping signs for both $(G_{N, \mbox{eff}} \to - G_{N, \mbox{eff}},~\Lambda_{4D, \mbox{eff}} \to - \Lambda_{4D, \mbox{eff}})$ keeps functional form of (\ref{efL_4}) and corresponds
to positive effective Newtonian constant as well as effective $4D$ $\Lambda$-term. Let us note that it is evident from (\ref{efL_5}) that even if $\Lambda = 0$ (globally vacuum solution, i.e. with no
boundary term), effective $4D$ $\Lambda$-term would be nonzero. The source for nonzero effective $4D$ $\Lambda$-term is the curvature of extra dimensions, and it could be (formally) both positive
and negative.

That is exactly what we meant when stating that $\Lambda$ in the original action (\ref{Sorig}) is a boundary term rather then cosmological term. Indeed, cosmological (or $\Lambda$-) term was introduced by
Einstein to make the Universe static, so that it relates to our $(3+1)$-dimensional world and it is $\Lambda_{4D, \mbox{eff}}$. On the contrary, original $\Lambda$ relates to the entire 
$(D+1)$-dimensional space-time and they relate as (\ref{efL_5}). These matters were discussed in more details in~\cite{CGPT1, CGPT2}. 

Finally, let us address stability of the solution. Similarly to the previously considered cases, we perturb original system around exact solution. Now the exact solution has very simple form,
which simplify functional form of the perturbed equations. Substituting $\dot H = H = 0$ as well as (\ref{efL_2}) into perturbed equation, one can rewrite them as

\begin{equation}
\begin{array}{l}
\dac{4D b_0^3}{D-3}\ddot\delta b(t) + \dac{8(2D-3)b_0^4}{(D-2)(D-3)}\dot\delta H(t) = 0, \\ \\
6\alpha b_0 (D-2) \dot\delta H(t) - (D-1)\delta b(t) = 0.
\end{array} \label{efL_7}
\end{equation}

The solution for $\delta b(t)$ is just harmonic function

\begin{equation}
\begin{array}{l}
\delta b(t) = C_1 \sin(\omega t) + C_2 \cos (\omega t)
\end{array} \label{efL_8}
\end{equation}

\noindent with

\begin{equation}
\begin{array}{l}
\omega^2 = \dac{(D-1)(2D-3)}{3D\alpha (D-2)^2}.
\end{array} \label{efL_9}
\end{equation}

\begin{figure}
\includegraphics[bb=28 125 575 663, width=0.5\textwidth, angle=0]{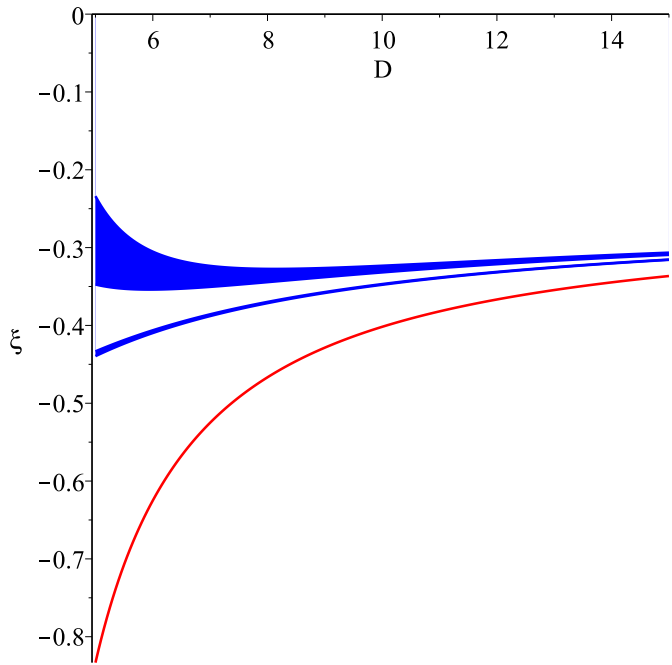}
\caption{Areas of stability on $(D, \xi)$ plane (in blue) and relation for $\xi$ from (\ref{efL_2}) (red) (see the text for more details).}\label{fig1_0}
\end{figure}

With a bit of math one can demonstrate that the result we reported previously (see Eq. (17) from~\cite{CGPT3}) is the same as (\ref{efL_9}), but this time it is derived for general $\gamma_D$. From
(\ref{efL_9}) one can see that the oscillating solution exist only for $\alpha > 0$, which corresponds to $\gamma_D < 0$. So that, despite the fact that we derived it for general case, oscillatory
solution exist only for $\gamma_D < 0$, which is described in detail in~\cite{CGPT3}). This also could be illustrated by Fig.~\ref{fig1_0}, where we plot areas of stability of solutions, derived
in~\cite{we2020} in blue and added value for $\xi$ from (\ref{efL_2}) as a red line. One can see that they do not intersect and this is ``graphical'' evidence that the oscillating regime does not
exist in this case.

\section{Conclusions}

Let us summarize our findings presented in current manuscript. First of all, we have proven that for the case of positive spatial curvature of extra dimensions
it is impossible to build two-stages scenario similar to the described in~\cite{CGT-2020}. Within this scenario,
on the first stage, initially totally anisotropic (Bianchi-I-like) $D+3$-dimensional
 space evolves to the state with 3-dimensional expanding and $D$-dimensional contracting isotropic subspaces; on the second stage, constant negative curvature of extra dimensions begins to play the 
 role and provide compactification of these extra dimensions. We have demonstrated that if the curvature of extra dimensions is positive, the areas of stability for the exponential (stage-1) 
 solution and for the asymptotically static extra dimensions (stage-2) do not overlap: depending on $\xi = \alpha\Lambda$ we could have either of them stable but not both, which destroy the scheme.
 
 In~\cite{we2020} we have provided the description of the general solution and its properties including lack of the solutions for specific signs of $\alpha$ and $\Lambda$ for some $D$ -- in the
 current manuscript with use of slightly different approach we demonstrated the same in a very elegant graphical way. 
 
 Finally, earlier in~\cite{CGPT3} we have demonstrated that for negative curvature of extra dimensions it is possible to recover Friedmann-like behavior with oscillations. Since in~\cite{we2020} we
 have demonstrated the possibility of building solution with positive curvature of extra dimensions, natural question arose -- is it possible to recover Friedmann-like late-time behavior for
 positive curvature as well? In this paper we investigated this question for generic curvature and found out that late-time Friedmann-like behavior could be obtained only for $\alpha > 0$ which
 corresponds to negative curvature. So that late-time Friedmann-like behavior is possible only in case of negative curvature.
 
 Overall, the results presented in this paper deepen the difference between the cases with negative and positive curvature. Previously in~\cite{we2020} we mentioned that the main difference between them 
 lies in the existence areas over parameters' space -- the case with negative curvature has it open, so that the probability to end up with that regime is quite high, while for the positive curvature
 the areas of stable existence are closed and the area itself is decreasing with growth of $D$. Analysis done in this paper brings two more ``disadvantages'' to the case with positive curvature --
 one cannot build two-stages scenario for this case (as the areas of stability for exponential and compactified solutions do not overlap) and one cannot restore ``nearly-Firedmann'' late-time
 regime. These facts suggest that, the case with positive spatial curvature of extra dimensions,
 despite being more ``aesthetically'' attractive (it is much easier to visualise positive spatial curvature; for the negative spatial curvature it is not easy
 to build topologically closed manifold, yet to define its ``size'' and visualize it), dynamically ``loses'' to the case with negative curvature.


\begin{thebibliography}{99}
\bibitem{Eddington} A. S. Eddington, The Mathematical Theory of Relativity, Cambridge: At the University Press (1923)

\bibitem{KK} T. Kaluza, 
Sitzungsber. Preuss. Akad. Wiss. Berlin (Math.Phys.), 1921:966–972 (1921);
O. Klein, 
Zeitschrift fur Physik, 37:895–906, (1926).

\bibitem{Weyl} H. Weyl, 
Sitzungsber. Preuss. Akad. d. Wiss. Teil, 1(465) (1918)

\bibitem{Mod-grav-cosmol} T. Clifton, P. G. Ferreira, A. Padilla and C. Skordis, 
Physics Reports 513, 1, 1-189 (2012)

\bibitem{Stelle} K. S. Stelle, 
Physical Review D 16(4), (1977)

\bibitem{Starobinsky}  A. A. Starobinsky, 
Physics Letters B, 91(1):99 – 102 (1980);
A. A. Starobinskii, 
Soviet Astronomy Letters, 9:302 (1983).

\bibitem{Sotiriou-Faraoni} T. Sotiriou and V. Faraoni, 
Reviews of Modern Physics, 82, 451–497 (2010)

\bibitem{Muller} D. Muller, A. Ricciardone, A. Starobinsky and A. Toporensky, 
Eur. Phys. J. C, 78, 311 (2018)

\bibitem{Lovelock} D. Lovelock, 
J. Math. Phys., 12, No. 3, 498–501 (1971)

\bibitem{add1} Demaret J., Caprasse H., Moussiaux A., Tombal P. and Papadopoulos D., 
Phys. Rev D 41, 1163 (1990);
Mena Marugán G. A., 
Phys. Rev. D 46, 4340 (1992);
Elizalde E., Makarenko A.N., Obukhov V.V., Osetrin K.E. and Filippov A.E., 
    Phys. Lett. B644, 1 (2007);
Maeda K.I. and Ohta N., 
JHEP 1406, 095 (2014);
Giribet G., Oliva J. and Troncoso R., 
JHEP 0605, 007 (2006);
Canfora F., Giacomini A., Troncoso R. and Willison S., 
Phys. Rev. D 80, 044029 (2009);
S.A. Pavluchenko, Phys. Rev. D {\bf 94}, 024046 (2016);
{\it ibid.} {\bf 94}, 084019 (2016);
S.A. Pavluchenko, Eur. Phys. J. C {\bf 77}, 503 (2017);
{\it ibid.} {\bf 78}, 551 (2018);
{\it ibid.} {\bf 78}, 611 (2018);
S.A. Pavluchenko, Particles {\bf 1}, 36 (2018)  [arXiv:1803.01887].

\bibitem{Muller-Hoissen-1} Muller-Hoissen F., 
Phys. Lett. B 163, 106 (1985)

\bibitem{Deruelle} Deruelle N. and Farina-Busto L., 
Phys. Rev. D 41, 3696 (1990)

\bibitem{Muller-Hoissen-2} Müller-Hoissen F., 
Class. Quant. Grav. 3, 665 (1986)

\bibitem{Ishihara} Ishihara H., 
Phys. Lett. B 179, 217 (1986)

\bibitem{Garraffo} C. Garraffo and G. Giribet, Mod. Phys. Lett.A23, 1801 (2008)

\bibitem{Ivashchuk} V.D. Ivashchuk, 
Int. J. Geom. Methods Mod. Phys. 7, 797 (2010)

\bibitem{DPT-15} D. Chirkov, S. Pavluchenko and A. Toporensky, 
Gen. Rel. Grav. 47, 137 (2015)

\bibitem{DPT-14} D.M. Chirkov, S.A. Pavluchenko and A.V. Toporensky, 
Mod. Phys. Lett. A29, 1450093 (2014)

\bibitem{CST2} D. Chirkov, S. Pavluchenko, A. Toporensky, Gen. Rel. Grav. {\bf 46}, 1799 (2014); \href{http://arxiv.org/abs/1403.4625}{arXiv:1403.4625}.


\bibitem{PK1} S.A. Pavluchenko and A.V. Toporensky, Mod. Phys. Lett. {\bf A24}, 513 (2009); I.V. Kirnos, A.N. Makarenko, S.A. Pavluchenko and A.V. Toporensky, Gen. Rel. Grav. 42, 2633
(2010).

\bibitem{PK2} S.A. Pavluchenko,   Phys. Rev. D {\bf 80}, 107501 (2009); I.V. Kirnos, S.A. Pavluchenko and A.V. Toporensky, Grav. Cosmol. 16, 274 (2010).


\bibitem{PT} S.A. Pavluchenko and A.V. Toporensky, Gravitation and Cosmology {\bf 20}, 127 (2014); \href{http://arxiv.org/abs/1212.1386}{arXiv:1212.1386}.

\bibitem{prd10} S.A. Pavluchenko,   Phys. Rev. D {\bf 82}, 104021 (2010).

\bibitem{my18d} S.A. Pavluchenko, Eur. Phys. J. C {\bf 79}, 111 (2019).



\bibitem{stab} S.A. Pavluchenko, 
Phys. Rev. D 92, 104017 (2015); 
V.D. Ivashchuk, Grav. Cosmol. 22(4), 329 (2016); 
D. M. Chirkov and A. V. Toporensky, Grav. Cosmol., 23(4), 359 (2017); 
K.K. Ernazarov and V.D. Ivashchuk, Eur. Phys. J. C 77, 402 (2017);
V. D. Ivashchuk and A. A. Kobtsev, Eur. Phys. J. {\bf C} 78, 100 (2018);
V. D. Ivashchuk, A. A. Kobtsev, arXiv:2009.10204.

\bibitem{Ivas-16} V.D. Ivashchuk, 
Eur. Phys. J.  C {\bf 76} 431 (2016)


\bibitem{infl} S.A. Pavluchenko, Phys. Rev. D {\bf 67}, 103518 (2003); {\it ibid.} {\bf 69}, 021301 (2004).



\bibitem{CGPT1}
F.~Canfora, A.~Giacomini and S.~A.~Pavluchenko,
  Phys.\ Rev.\ D {\bf 88}, no. 6, 064044 (2013)

\bibitem{CGPT2}
F.~Canfora, A.~Giacomini and S.~A.~Pavluchenko,
  Gen.\ Rel.\ Grav.\  {\bf 46}, no. 10, 1805 (2014)

\bibitem{CGPT3} F. Canfora, A. Giacomini, S. A. Pavluchenko and A. Toporensky, 
Grav. Cosmol. 24, 28 (2018)

\bibitem{CGT-2020}
D. Chirkov, A. Giacomini, and A. Toporensky, 
Gen. Rel. Grav. {\bf 52}, 30 (2020).

\bibitem{PT2017} S.A. Pavluchenko and A.V. Toporensky, Eur. Phys. J. C {\bf 78}, 373 (2018).

\bibitem{we2020}
D. Chirkov, A. Giacomini,  S.A. Pavluchenko and A. Toporensky, 
arXiv:2012.03517



\end{thebibliography}
\end{document}